\documentclass{PoS}
\usepackage{epsfig}
\usepackage{graphicx}
\newcommand\KK{{\cal KK}}
\title{HERWIRI2.1: Electroweak Corrections for Hadron Scattering}
\ShortTitle{HERWIRI2.1: Electroweak Corrections for Hadron Scattering}
\author{\speaker{Scott Yost}%
         \thanks{This work and its presentation were supported in part by grants from The Citadel Foundation.}\\
        The Citadel\\
        Charleston, SC 29409, USA\\
        and\\
        Institute for Nuclear Physics, Polish Academy of Science (IFJ-PAN)\\
        Krak{\'{o}}w, Poland\\
        E-mail: \email{scott.yost@citadel.edu}}
\author{B.F.L.\ Ward\\
        Baylor University\\
        Waco, TX 76798, USA\\
        E-mail: \email{bfl\_ward@baylor.edu}}
\abstract{
We describe the program HERWIRI2.1, which implements ${\cal O}(\alpha^2)$ 
photonic radiative corrections exponentiated at the amplitude level (initial 
state, final state, and initial-final interference) and electroweak corrections
to the matrix element by generating events using the \KK\ Monte Carlo to 
generate the hard process, with quark initial states generated according to 
PDFs via an LHAPDF interface. The events can be showered internally using 
HERWIG6.5 or exported and showered using any desired external showering 
program. Some early results are shown, including comparisons to HERWIG6.5 
and HORACE 3.1.

\vspace{1cm}
\begin{center}
\hbox{IFJPAN-IV-2016-16\newline
BU-HEPP-16-03}
\end{center}
}

\FullConference{Loops and Legs in Quantum Field Theory\\
		 24-29 April 2016\\
		 Leipzig, Germany\\[1cm] 
                 
		 {\rm IFJPAN-IV-2016-16}\\
		 {\rm BU-HEPP-16-03}}

\begin{document}
\section{Introduction}
Vector boson production plays an important role at the LHC, both as a benchmark
process for precision tests of the Standard Model and as a background in 
searches for new physics.\cite{LHCPrecision}  It is desired to measure this 
process at the 1\% level, but electroweak corrections alone are typically 
several percent, as seen in the studies of Ref.\ \cite{EWsys1,EWsys2}.
A number of programs are presently available for calculating electroweak 
corrections to hadron scattering, and a comprehensive review of the state of 
the art in calculating vector boson observables at the LHC can be found in 
Ref.\ \cite{LHCPrecision}, which presents tuned comparisons using 
HORACE \cite{Horace1,Horace2}, PHOTOS \cite{Photos1,Photos2,Photos3}, 
SANC \cite{SANC}, WINHAC \cite{WINHAC1,WINHAC2,WINHAC3}, 
WZGRAD \cite{WZGRAD1,WZGRAD2,WZGRAD3}, and FEWZ3 \cite{FEWZ3}, together with 
programs implementing higher order QCD effects, which are not presently 
included in HERWIRI2.1, but will be implemented in a future version.


The name HERWIRI is an acronym for {\bf H}igh {\bf E}nergy {\bf R}adiation 
{\bf W}ith {\bf I}nfra-{\bf R}ed {\bf I}mprovements, and recognizes the role 
HERWIG \cite{HERWIG1,HERWIG2,HERWIG3} has in the project. HERWIRI2.x, which 
concentrates on electroweak and exponentiated photonic corrections, is 
independent of HERWIRI1.x \cite{HERWIRI1-1,HERWIRI1-2}, 
another project under the HERWIRI umbrella which implements a non-abelian 
analog of the the amplitude-level YFS-inspired \cite{YFS} exponentiation 
developed for \KK MC \cite{KKMC} to modify the soft behavior of the parton 
shower. 

HERWIRI2.1 adapts \KK MC, which was developed as a high-precision 
generator for $e^+ e^-\rightarrow\gamma,Z\rightarrow f{\overline f}$ for LEP
physics, to hadronic collisions. An early, unreleased version of HERWIRI2
\cite{HERWIRI2} was based on the original KKMC and a reweighting 
scheme. However, \KK MC was later upgraded to version 4.22 \cite{KKMC422}, 
which supports quark initial states, and this version became the basis of
 HERWIRI2.1, which generates quark initial states using 
an LHAPDF \cite{LHAPDF} interface integrated into \KK MC 4.22. 

HERWIRI2 generates a hard process for $pp$ or $p{\overline p}$ scattering
including ISR and FSR photons added via \KK MC, and the events may be exported
for external showering, or showered internally using HERWIG6.5, which is 
included in any case as a means of incorporating HEPEVT bookkeeping and PDF
management in a form which is consistent with HERWIG. Tau lepton events may
be passed to a shower as-is, or tau decay may be implemented via TAUOLA
\cite{TAUOLA1,TAUOLA2,TAUOLA3} and PHOTOS, which are incorporated into \KK MC. 

A HERWIRI2.1 run begins with an initialization phase when parameters are set,
both in \KK MC and HERWIG6.5. This insures compatible parameters if the internal
HERWIG shower is used. Even without an internal shower, HERWIG routines are 
used for HEPEVT and PDF management. In addition, the underlying MC generator,
FOAM, creates an adaptive grid for the subsequent event generation and 
calculates a crude integral, as explained in Section 3.

\section{EEX and CEEX Exclusive Exponentiated Cross Sections}
HERWIRI2.1 incorporates \KK MC's options for calculating exponentiated ISR and
FSR using either cross-section level exponentiation (EEX) or amplitude-level
exclusive exponentiation (CEEX).  The default CEEX mode includes initial-final
interference (IFI) effects and exact emission to order $\alpha$, $\alpha^2 L^2$,
and $\alpha^2 L$,  where $L = \ln(p^2/m^2)$ with the hard process scale $p^2$ 
and the relevant mass $m$, which is the final fermion mass for FSR and quark 
current mass for ISR. This is referred to as 
${\cal O}(\alpha^2)$-pragmatic.\cite{KKMC}  EEX mode includes exact order 
$\alpha$, as well as exact order $\alpha^2L^2$, $\alpha^2L$, and 
$\alpha^3 L^3$, which is referred to as ${\cal O}(\alpha^3)$-pragmatic. 
The most precise ${\cal O}(\alpha^2)$ corrections in KKMC may be found in
Ref.\ \cite{exact-brem}.
Electroweak boson mixing enters at the $\alpha^2$ non-logarithmic order, 
so EEX includes most-precise ${\cal O}(\alpha^2)$ corrections for which full 
electroweak gauge invariance can be maintained in QED alone. When run in CEEX 
mode, alternative weights are provided for EEX and lower orders in $\alpha$ as 
well. There is also a CEEX weight without IFI, for comparison.

The cross section takes the form 
\begin{equation}
\sigma = \sum_i \int {dx_1} {dx_2} dv f_i(p^2, x_1) 
f_{\overline i}(p^2, x_2) \sigma(p^2, v) \delta(p^2 - sx_1 x_2)
\end{equation}
where $p^2 = (p_1 + p_2)^2$ for initial quark momenta $p_1$ and $p_2$ given as
fractions $p_j = x_j P_j$ of the proton or anti-proton beam momenta $P_j$ ($j$
= 1,2), $i$ indexes all relevant quark and antiquark flavors, and $f_i$ are the
parton distribution functions. In the EEX case, the squared invariant mass 
after ISR is $(p - K_I)^2 = p^2(1 - v)$, where $K_I$ is the total initial 
photon momentum, and
\begin{eqnarray}
\sigma(p^2, v) &=& \sum_{n=0}^\infty \sum_{n'=0}^\infty 
\frac{1}{n!n'!}  \exp(Y_I(\Omega_I; p^2)) \int Dq_1 Dq_2 
\exp(Y_F(\Omega_F; q^2)) \nonumber\\
& &\int \prod_{j=1}^{n}Dk_j (1-\Theta(\Omega_I;k_j))
\int \prod_{j'=1}^{n'} Dk'_{j'}  
	(1-\Theta(\Omega_F;k_{j'}))
\delta\left(K_I -\sum_{j=1}^n k_j\right) 
\label{sigma-EEX}
\\
& & 
\delta\left(v - \frac{(2p-K_I)\cdot K_I}{p^2}\right)
\rho_{\rm EEX}^{(n,n')}
(p_1, p_2; q_1, q_2; k_1,\ldots, k_n; k'_1,\ldots, k'_{n'})
\nonumber
\end{eqnarray}
Here, $Dk$ represents the Lorentz invariant phase space density for momentum 
$k$, $\Omega_I$ and 
$\Omega_F$ are soft photon regions for ISR and FSR, $\Theta(\Omega, k) = 1$ or
0 depending on whether $k\in\Omega$ or not, ${\widetilde S}_{I,F}$ are 
initial and final state soft photon factors, and $Y_{I,F}$ are initial and
final state YFS factors. The density $\rho_{\rm EEX}^{n,n'}$ with $n$ ISR and
$n'$ FSR photons is constructed from a combination of soft photon factors 
${\widetilde S}(k)$ and $N$-photon YFS residuals 
${\overline\beta}_N(k_1,\ldots k_N)$, 
\begin{equation}
\rho_{\rm EEX}^{(n,n')} = \prod_{j=1}^{n}{\widetilde S}_j
\prod_{j'=1}^{n'}{\widetilde S}_{j'} \left\{ {\overline\beta}_0 + \sum_{j=1}^n 
\frac{{\overline\beta}_1(k_j)}{{\widetilde S}_I(k_j)} + \sum_{j = 1}^{n'}
\frac{{\overline\beta}_1(k_{j'})}{{\widetilde S}_F(k_j)} + \ldots\right\}. 
\end{equation}
Detailed expressions may be found in Ref.\ \cite{KKMC},
with the incoming electron replaced by a quark.  The YFS residuals are all 
evaluated at a scale $s_X \equiv p^2(1 - v)$, which represents the scale of the
hard $\gamma/Z$ process.

In the CEEX case, the photons cannot be sorted unambiguously into ISR and 
FSR subsets, so the sum over all such possibilities is performed, and $v$ 
plays an analogous role for the sum over ISR photons in a given partition. 
The analog of (\ref{sigma-EEX}) is
\begin{eqnarray}
 \sigma_{\rm CEEX}(p^2, v) &=& \sum_{N=0}^\infty \sum_{\cal P}
\frac{1}{N!}  \int Dq_1 Dq_2 
\exp(Y_F(\Omega; p_1, p_2, q_1, q_2)) 
 \int \prod_{j=1}^{N}Dk_j (1-\Theta(\Omega;k_j))\nonumber\\
& & \hspace{-1in}
\delta\left(v - \frac{(2p-K_I({\cal P}))\cdot K_I({\cal P})}{p^2}\right)
\delta\left(K_I({\cal P}) -\sum_{j=1}^{n({\cal P})} k_j\right) 
\rho_{\rm CEEX}^{(N, {\cal P})} (p_1, p_2; q_1, q_2; k_1,\ldots, k_N),
\label{sigma-CEEX}
\end{eqnarray}
where ${\cal P}$ is a partition of $N$ photons into ISR and FSR subsets, which
are summed in all possible ways with $n({\cal P})$ ISR photons and 
$n'({\cal P}) = N - n({\cal P})$ FSR photons, and 
$\rho^{(N, {\cal P})}_{\rm CEEX}$ is constructed as a sum of absolute squares 
of helicity amplitudes
\begin{equation}
{\cal M}_N^{\cal P} = \prod_{i = 1}^{n(\cal P)} {\cal S}_I(k_i) 
\prod_{i'=1}^{n'(\cal P)} {\cal S}_F(k_{i'}) \left\{{\widehat\beta}_0 
+ \sum_{j=1}^n \frac{{\widehat\beta}_1(k_j)}{{\cal S}_I(k_j)}
+ \sum_{j'=1}^{n'} \frac{{\widehat\beta}_1(k_{j'})}{{\cal S}_I(k_{j'})}
+ \ldots\right\}, 
\end{equation}
where the ${\widehat\beta}_N$ are spinor-amplitude residuals and
the absolute squares of the soft-photon spinors ${\cal S}_{I}$ and 
${\cal S}_{F}$ yield ${\widetilde S}_I$ and ${\widetilde S}_F$, respectively,
while the cross terms are responsible for IFI corrections. A common soft-photon
set $\Omega$ is used for both ISR and FSR, since the assignment to these sets
is arbitrary. Again, the detailed
forms may be found in Ref.\ \cite{KKMC}, with incoming quarks replacing
electrons. 

\section{Primary Distribution}

The cross section is constructed as a Monte Carlo integral over 
$p \equiv \sqrt{p^2}$, $x_1$, $v$ and a sum over quark and antiquark flavors. 
The underlying FOAM MC generates
these according to a crude primary distribution which it integrates during the 
event generation.  The C++ version of FOAM is used, 
which allows the quark flavor dimension to be treated separately, with a 
predefined separation into five flavors and five anti-flavors, and no cell 
division.  The remaining three dimensions are divided into simplicial cells
during the exploration phase. 

Specifically, the default mappings used are 
\begin{equation}
p^{-2} = \frac{1 - r_1}{p^2_{\rm min}} + \frac{r_1}{p^2_{\rm max}}, 
\qquad 
x_1 = \left(\frac{p^2}{s}\right)^{r_2}, \qquad
v = v_{\rm max} r_3^{1/\gamma}, 
\end{equation}
for $0\le r_i < 1$, where $\gamma = {\overline\gamma} - 2Q_i^2\alpha/\pi$ and
\begin{equation}
\overline\gamma = \frac{2Q_i^2\alpha}{\pi} \ln\left(\frac{(1+\beta)^2}{m_i^2}\right)
\end{equation}
for a quark with charge $Q_i$ and current mass $m_i$, 
with $\beta = \sqrt{1 - 4m_i^2/s}$. 

The FOAM integrand for the crude MC integral is
\begin{eqnarray}
\rho_i(p^2, v) &=& 2N_q p^4 (p^{-2}_{\rm max} - p^{-2}_{\rm min})\ln\left(
    \frac{p^2}{s} \right) 
    \frac{\overline\gamma}{2\gamma} \left(1 + (1-v)^{-1/2}\right) 
    v_{\rm max}^{\gamma}
    \left(\frac{v}{v_{\rm min}}\right)^{2Q_i^2\alpha/\pi}\nonumber\\
& & f_i(p^2, x_1) f_{\overline i}(p^2, x_2) \sigma_{\rm Born}(p^2(1-v)),
\end{eqnarray}
with quark or antiquark label $i$ chosen by dividing the unit interval into
$2N_q$ equal intervals and selecting $i$ according to the box into which the
first random variable falls. 

\section{Electroweak Matrix Element Corrections}
Electroweak matrix element corrections are added using version 6.21 of the 
DIZET \cite{DIZET} package developed for ZFITTER \cite{ZFITTER}. The $\gamma$ 
and $Z$ propagators are multiplied by vacuum polarization factors of the form
\begin{equation}
{H_\gamma} = \frac{1}{2-\Pi_\gamma}, \qquad
{H_Z} = 4 \sin^2(2\theta_{\rm W}) \frac{\rho_{\rm EW} G_\mu M_Z^2}{8\pi \alpha 
        \sqrt{2}}.
\end{equation}
Vertex corrections are incorporated into the coupling of $Z$ to $f$ via
form factors in the vector coupling:
\begin{equation}
g_V^{(Z,f)} = \frac{T_3^{(f)}}{\sin(2\theta_W)} - Q_f {F_{\rm v}^{(f)}(s)} 
	\tan\theta_W.
\end{equation}
Box diagrams contain these plus a new angle-dependent form-factor in the 
doubly-vector component:
\begin{equation}
g_V^{(Z,i)} g_V^{(Z,f)} = 
\frac{1}{\sin^2(2\theta_W)}
\left(T_3^{(i)} T_3^{(f)} - 2T_3^{(i)} Q_f F_{\rm v}^{(f)}(s)
 - 2Q_i T_3^{(f)} F_{\rm v}^{(i)}(s) + 4Q_i Q_f F_{\rm box}^{(i,f)} (s,t)
\right) .
\end{equation}
The quantities in these expressions are tabulated at the beginning of the 
run for hard process energies up to 1040 GeV, the highest energy supported by 
the \KK MC interface to DIZET. When EW corrections are required for higher 
energies, they are calculated as needed.

\section{Results and Comparisons}
In this section, we present results both with and without a hadron shower for
proton collisions with a CM energy of 8 TeV,
using a cut 50 GeV $< p <$ 200 GeV on the generated quark scale 
$p = \sqrt{p^2}$.  In these tests, only muons are generated. 
The IR boundary is defined by setting a minimum value 
$v_{\rm min} = 10^{-3}$ in (\ref{sigma-EEX}) or (\ref{sigma-CEEX}). In 
principle, $v_{\rm min}$ could be set to $10^{-5}$ or less, but smaller values 
lead to lower efficiency. Electroweak parameters are taken from the benchmark 
choices in Ref.\ \cite{LHCPrecision}. HERWIRI2 was run in variable-width mode
for these tests. Comparisons are made to HERWIG6.5 and HORACE3.1, which use
fixed widths, so the $Z$ mass and width parameters were redefined in those 
programs as suggested by Ref.\ \cite{LHCPrecision} to maximize compatibility. 
All plots are for 25 million event runs with variable weights. 

Figs.\ \ref{fig:shower1} -- \ref{fig:shower3} show results for a run showered
with HERWIG6.5. The choice of shower was made largely for simplicity, since
HERWIG6.5 is already used for some PDF support and bookkeeping functions in
HERWIRI2.1, but also because it was the basis for the studies of 
Refs.\ \cite{EWsys1} and \cite{EWsys2}. HERWIRI2.1 results are shown for three
cases: the best full CEEX and EEX results with ISR and FSR (red), the 
CEEX result restricted to FSR only (blue), and a result with no photons 
(green) generated by HERWIRI2.1 using HERWIG's hard cross section. 
These are compared to 
HERWIG's result (black). The cross sections in each case are shown in
Table \ref{tab:showered}.  The best EEX result agrees with the CEEX cross
section to 0.04\%. A comparison is also shown for HORACE3.1, with the same
parameters. The HORACE comparison should be considered preliminary until it
has been more carefully tuned, but the agreement of the signs and general 
magnitude appear promising.

\begin{table}[h]
\begin{centering}
\begin{tabular}{|l|l|l|c|}
\hline
{\bf\ Generator} & {\bf\ Photons} & {\bf\ \ Cross Section} & {\bf \% Dif.}\\
\hline
HERWIG6.5  & None & $1039.6\pm 0.2$ pb& $-$\\
HERWIRI2.1 & CEEX ISR+FSR & $986.21\pm 0.26$ pb &$-5.1$\% \\
HERWIRI2.1 & EEX ISR+FSR & $985.82\pm 0.26$ pb &$-5.2$\% \\
HERWIRI2.1 & CEEX FSR & $986.05\pm 0.11$ pb &$-5.2$\%\\
HERWIRI2.1 & None (HERWIG Born)& $1038.69\pm 0.08$ pb &$-0.09$\%\\ 
HORACE3.1  & FSR & $1009.15\pm 0.79$ pb &$-2.9$\%\\
\hline
\end{tabular}
\caption{Cross sections for the showered tests. All results are for 25 million
events. Differences are relative to HERWIG6.5.}
\label{tab:showered}
\end{centering}
\end{table} 

Figures 1 - 3 show a collection of results for 25-million event runs of 
HERWIRI2.1 in CEEX mode
(${\cal O}(\alpha^2)$-pragmatic), both the full ISR+FSR result (which includes
IFI as well) in red and FSR alone in blue, showered with HERWIG6.5. 
The result of HERWIG6.5 alone is shown for comparison. Figures 4 - 6 show
results of 25-million event runs without a shower, in which case comparisons to
the HERWIG6.5 hard process (in black) and to HORACE3.1 (in green) are shown. 
While the cuts are different from the studies in Refs.\ \cite{EWsys1} and 
\cite{EWsys2}, the sign and general order of magnitude are as anticipated.

\FIGURE[ht]{
\label{fig:shower1}
\setlength{\unitlength}{1in}
\begin{picture}(6.5,2.4)(0,0)
\put(0,0.2){\epsfig{file=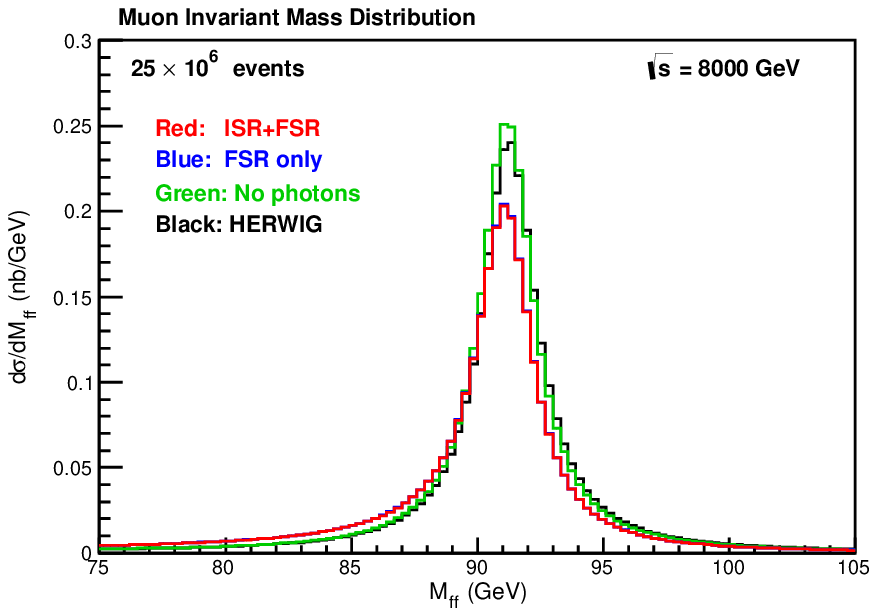,width=3.0in}}
\put(3,0.2){\epsfig{file=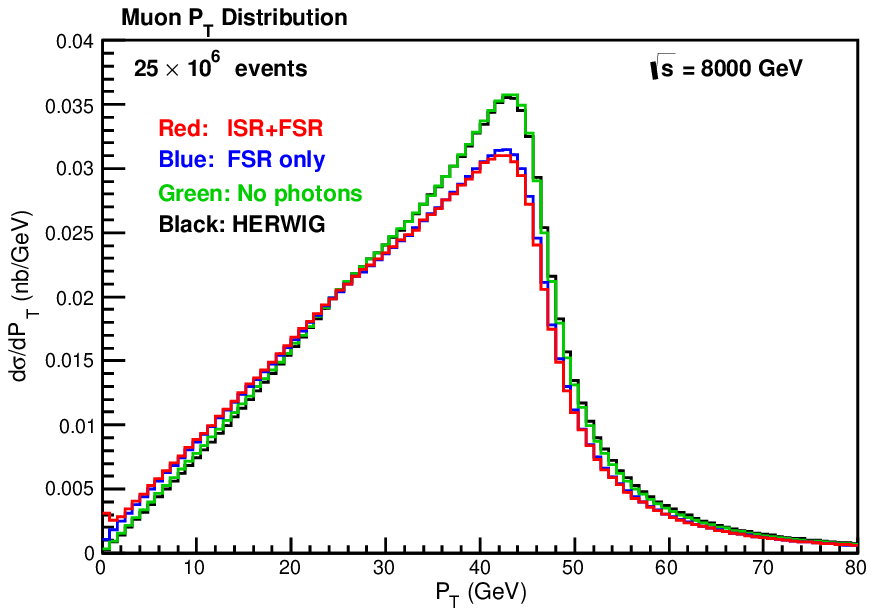,width=3.0in}}
\end{picture}
\vspace{-1cm}
\caption{Muon invariant mass and transverse momentum distributions
for HERWIRI2.1 for ISR + FSR (red) and for FSR alone (blue), 
showered by HERWIG6.5. The green distribution is generated by HERWIRI2.1 
using HERWIG's hard cross section and the black distribution is made by
HERWIG6.5 alone.}
}

\FIGURE[ht]{
\label{fig:shower2}
\setlength{\unitlength}{1in}
\begin{picture}(6.5,2.2)(0,0)
\put(0,0.2){\epsfig{file=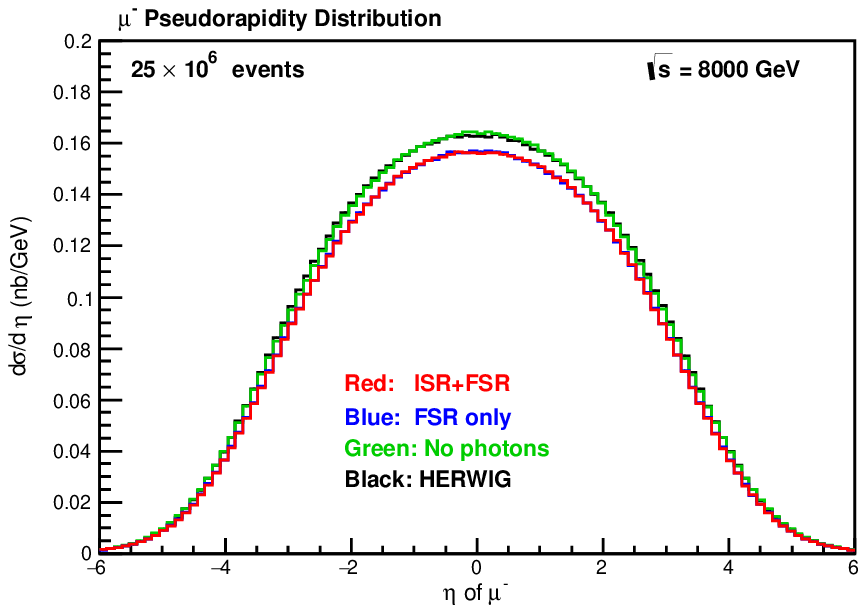,width=3.0in}}
\put(3,0.2){\epsfig{file=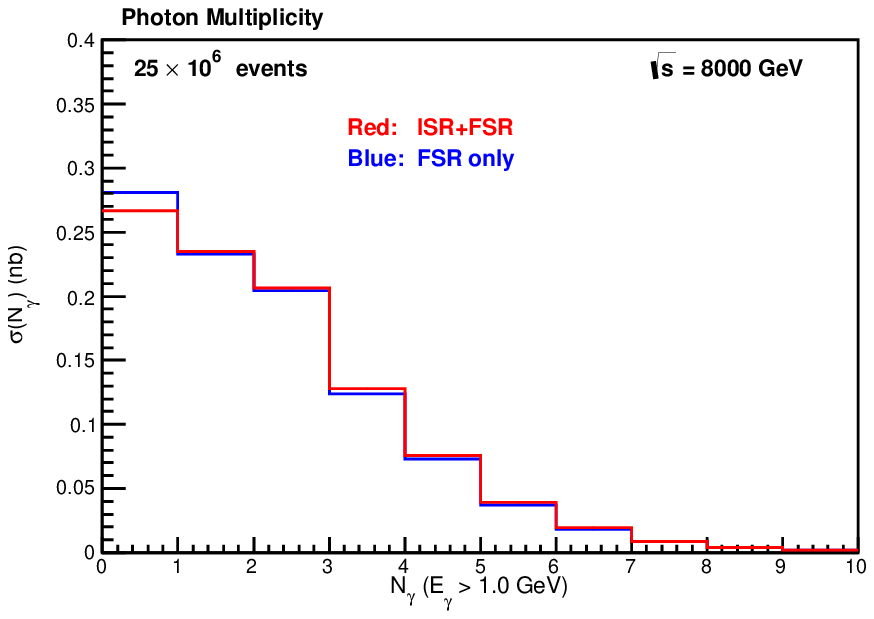,width=3.0in}}
\end{picture}
\vspace{-1.2cm}
\caption{Muon pseudorapidity distribution plotted as in the previous figure,
and the number of photons having energy greater than 1 GeV  for HERWIRI2.1 
with ISR + FSR (red) and FSR alone (blue), showered by HERWIG6.5.}
}

\FIGURE[ht]{
\label{fig:shower3}
\setlength{\unitlength}{1in}
\begin{picture}(6.5,2.2)(0,0)
\put(0,0.2){\epsfig{file=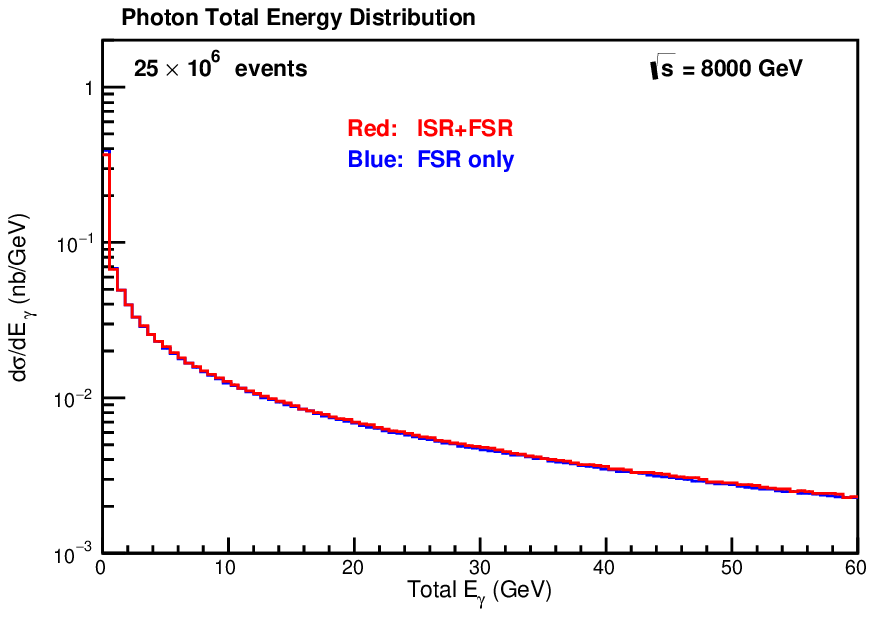,width=3.0in}}
\put(3,0.2){\epsfig{file=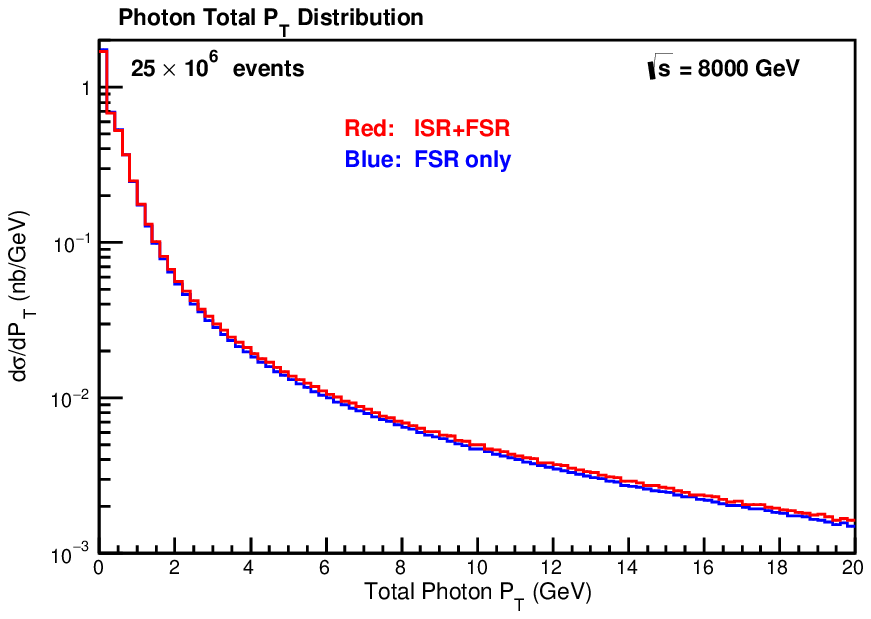,width=3.0in}}
\end{picture}
\vspace{-1.2cm}
\caption{Photon total energy distribution  and total photon transverse
momentum for HERWIRI2.1 showered by HERWIG6.5, plotted as in the previous 
figure.}
}

\FIGURE[ht]{
\label{fig:noshower1}
\setlength{\unitlength}{1in}
\begin{picture}(6.5,2.0)(0,0)
\put(0,0.2){\epsfig{file=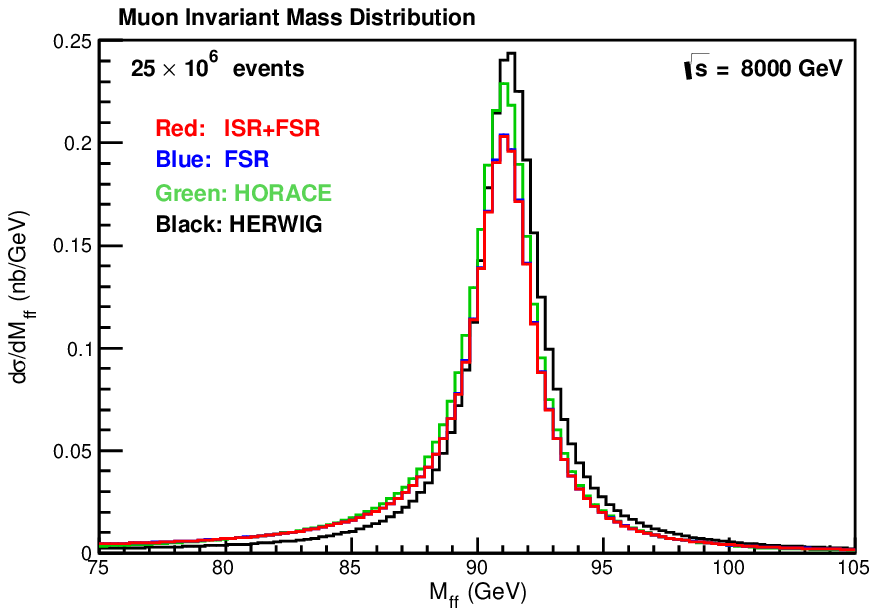,width=3.0in}}
\put(3,0.2){\epsfig{file=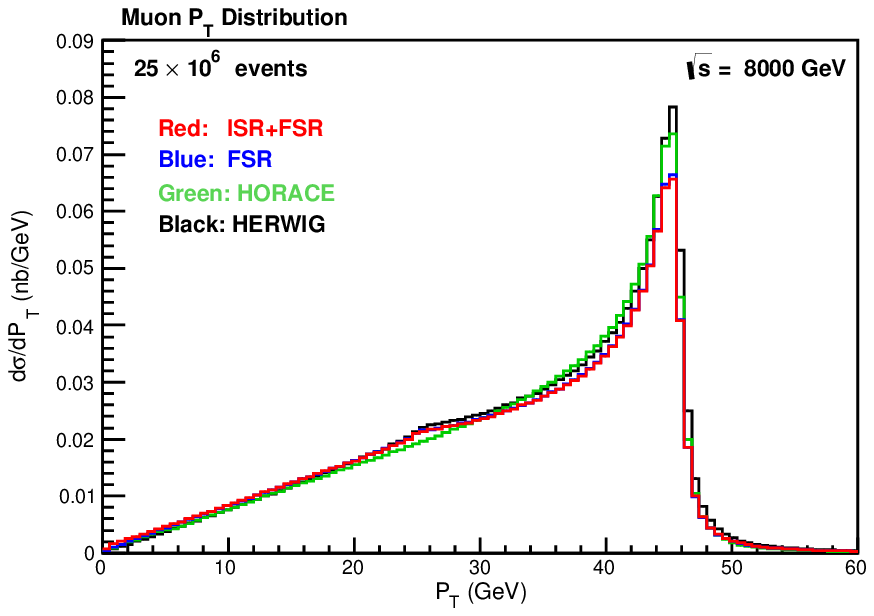,width=3.0in}}
\end{picture}
\vspace{-1.2cm}
\caption{Muon invariant mass and transverse momentum distributions
for HERWIRI2.1 with ISR + FSR (red) and FSR only (blue),
compared to HORACE3.1 (green) and the unshowered HERWIG6.5 hard
cross section (black). }
}

\FIGURE[ht]{
\label{fig:noshower2}
\setlength{\unitlength}{1in}
\begin{picture}(6.5,2.2)(0,0)
\put(0,0.2){\epsfig{file=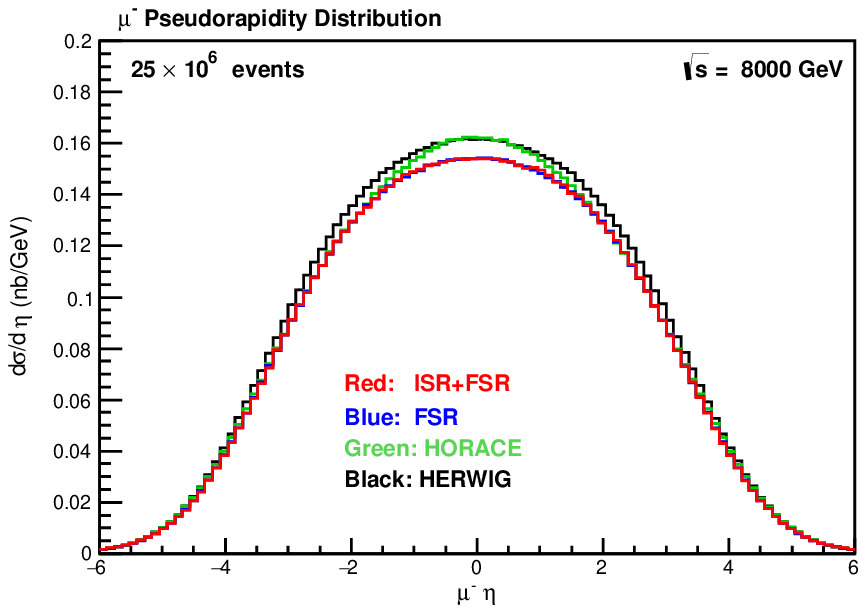,width=3.0in}}
\put(3,0.2){\epsfig{file=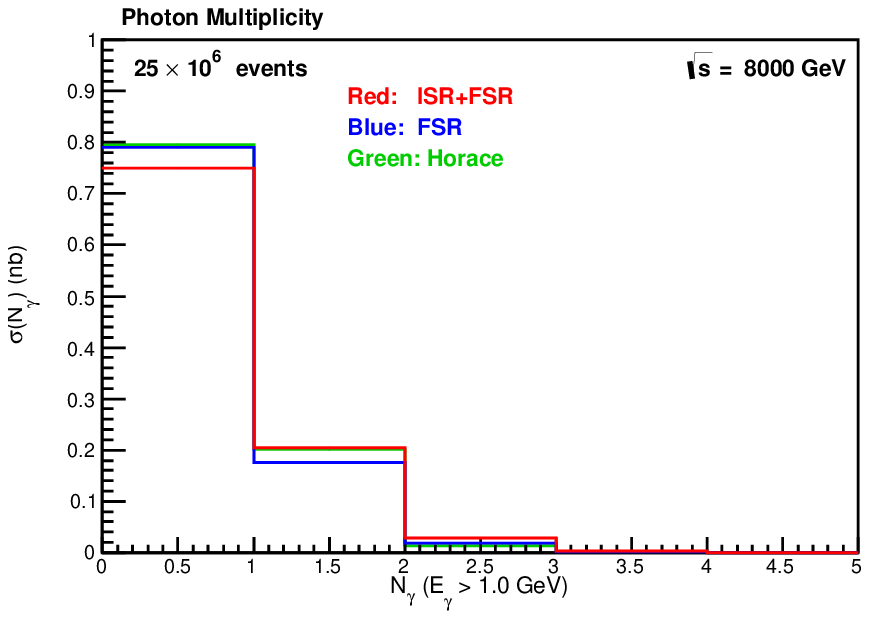,width=3.0in}}
\end{picture}
\vspace{-1.2cm}
\caption{Muon pseudorapidity distribution plotted as in the previous figure
and the number of photons with energy greater than 1 GeV 
for HERWIRI2.1 with ISR + FSR (red) and FSR only (blue), 
compared to HORACE3.1 (green).}
}

\FIGURE[ht]{
\label{fig:noshower3}
\setlength{\unitlength}{1in}
\begin{picture}(6.5,2.0)(0,0)
\put(0,0.2){\epsfig{file=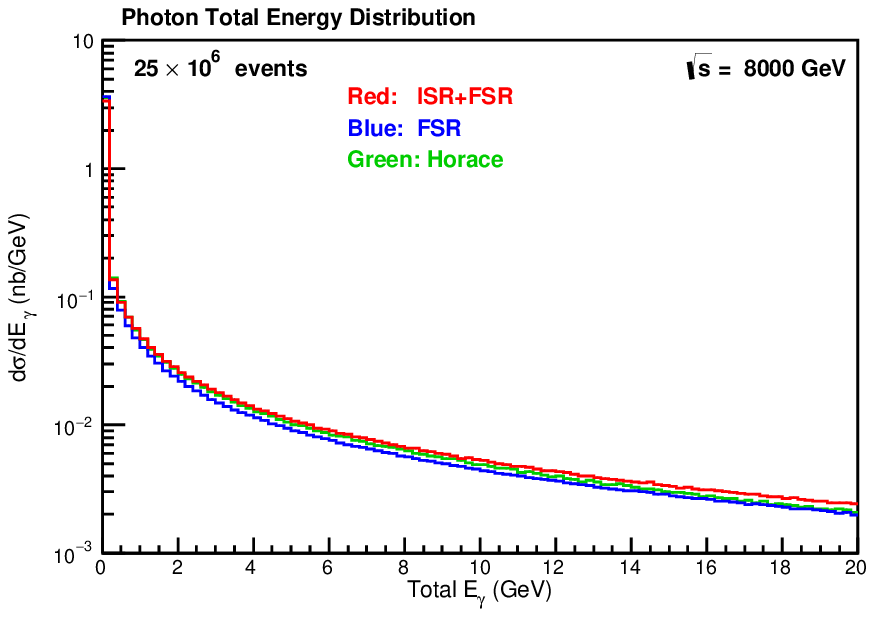,width=3.0in}}
\put(3,0.2){\epsfig{file=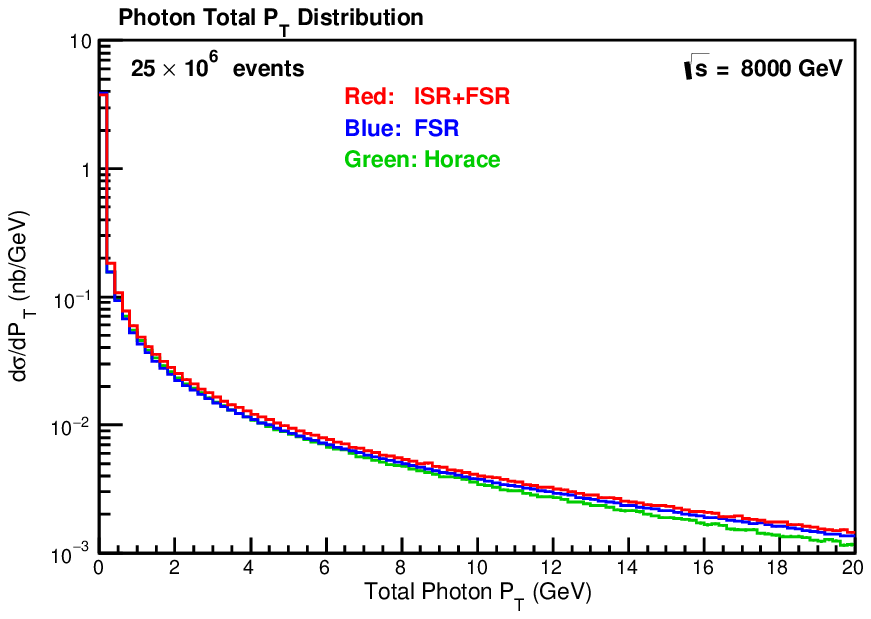,width=3.0in}}
\end{picture}
\vspace{-1.2cm}
\caption{Total photon energy and total photon transverse momentum 
for HERWIRI2.1 compared to HORACE3.1 as in the previous figure.}
}

\section{Conclusions and Outlook}

We have described the event generator HERWIRI2.1 for $pp$ or $p{\overline p}$
collisions with order $\alpha$ electroweak matrix element corrections and 
order $\alpha^2$-pragmatic CEEX photon emission (ISR, FSR, and IFI), and 
presented some of its first results, including comparisons to HERWIG and 
HORACE. The early results are promising and generally in accord with 
expectations. Further comparisons with more careful tuning and a variety of 
final fermions and cuts will be forthcoming, as will showering with a 
variety of modern generators.  We also intend to add NLO QCD corrections in 
a factorized approximation. 

\section*{Acknowledgments}
We acknowledge the hospitality of the CERN theory division, which contributed
greatly to the completion of HERWIRI2.1. S. Yost acknowledges the hospitality
and support of the Theoretical Physics Division of the Institute for Nuclear
Physics of the Polish Academy of Science and a sabbatical funded by The 
Citadel Foundation. S. Yost also acknowledges hospitality and support from 
V. Halyo, D. Marlow, and Princeton University during development of HERWIRI2.0,
which also received support from a U.S. Department of Energy grant.

\end{document}